Ferroelectricity in crystals with non-polar point groups

Yuxuan Sheng[1], Menghao Wu[1,2†]

[1]School of Physics, Huazhong University of Science and Technology, Wuhan, China 430074

[2]School of Chemistry, Institute of Theoretical Chemistry, Huazhong University of Science and Technology, Wuhan, China 430074

†wmh1987@hust.edu.cn

Abstract    Ferroelectric crystals must adopt one of the 10 polar point groups according to the Neumann's principle. In this paper we propose that this conclusion is based on perfect bulk crystals without taking the boundaries into account, and we show first-principles evidence that ferroelectric polarizations may also be formed in some non-polar point groups as the edges generally break the crystal symmetry, which may even maintain at macroscale. They can be switchable in some systems with weak van der Waals bondings or covalent-like ionic bondings where long ion displacements with moderate barriers are possible. Such unconventional ferroelectricity violates the Neumann's principle and Abrahams' conditions respectively due to the boundaries and long ion displacements, which may explain some unclarified phenomena reported previously as well as significantly expand the scope of ferroelectrics.

Ferroelectric materials with hysteretic electrically switchable polarizations have been widely studied in the past century, which have also impacted various applications ranging from actuators to data storage.[1] The spontaneous polarization of a ferroelectric crystal is associated with the permanent electric dipole moment in a unit cell. For a molecule, its dipole moment points from the center of the negative charge distribution to the center of the positive one, and the molecule should possess a permanent dipole moment if the two centers do not coincide. However, the polarization of a solid crystal cannot be simply estimated by using the formula $\Delta P = \frac{1}{\Omega} \int r \delta(r) d^3 r$ that depends on the shape of the chosen unitcell. 30 years ago, King-Smith and Vanderbilt revealed that upon a finite adiabatic change in the Hamiltonian of a system, the electronic contribution to the difference in polarization can be identified as a geometric quantum phase or Berry phase of the valence wave functions, and the polarization is only well-defined modulo $fe\mathbf{R}/\Omega_0$, where **R** is a lattice vector.[2] According to the Neumann's principle[3], the symmetry operations of any physical property of a crystal must include the symmetry operations of the point group of the crystal. The absence of polarization in 11 centrosymmetric point groups and 11 noncentrosymmetric-nonpolar point group can be deduced, and ferroelectrics are supposed to belong to the 10 polar point groups with an unique rotation axis, no inversion center and no mirror plane [4,5]: *$C_1$, $C_2$, $C_s$, $C_{2v}$, $C_3$, $C_{3v}$, $C_4$, $C_{4v}$, $C_6$, $C_{6v}$*., despite very few theoretical [6-8] and experimental reports [9] that have already violated the Neumann's principle[8]. The symmetry group of ferroelectric phase should be a maximal subgroup of high-temperature paraelectric symmetry group along the polarization direction, and based on group-to-subgroup relationship, there are 88 species of paraelectric-to-ferroelectric phase transitions given by Aizu.[10]

We note that those conclusions are based on perfect crystals without taking the boundaries into account, while the edges may break the crystal symmetry. The number of surface atoms in a bulk structure are negligible compared with the atoms inside, so in general, the bulk properties are scarcely dependent on surface. In this paper, we predict the existence of unconventional ferroelectricity in some crystals with nonpolar point groups when the boundaries are taken into consideration. As a paradigmatic case, for a two-dimensional (2D) binary honeycomb monolayer (e.g., BN monolayer) in Fig. 1(a) composed with A cations with charge δ and B anions with charge −δ, its point group with $D_{3h}$ symmetry is non-polar. The corresponding equilateral hexagonal or triangular flakes (no matter how large their sizes are, see Fig. 1(b)) are also non-polar since the center of positive charge and negative charge coincide.

However, for a rectangular flake with two zigzag edges and two armchair edges, the boundaries do not match the crystal symmetry. Despite there are 3 identical armchair directions in an infinite honeycomb lattice, the boundaries of the rectangular flake lead to inequivalence between them, and an in-plane polarization will be formed parallel to the armchair edges. For a rectangular flake composed integer number of rectangle unitcells, as shown in Fig.1(c), without edge reconstructions ideally, the total dipole moment is estimated to be nmδd by the summation of $\Delta P = \frac{1}{\Omega} \int r \delta(r) d^3 r$, where n, m, and d denote the number of units along the zigzag and armchair directions, and A-B bond length, respectively. Over the total area $3\sqrt{3}nmd^2$, the average areal polarization of $\delta/3\sqrt{3}d$ is constant and does not depend on the size of rectangular flake. It seems that the edges may determine the ferroelectric polarizations even at macroscopic scale. The actual magnitude measured in experiment can be quite different since such polarization is highly dependent on boundaries and may be greatly reduced by edge reconstruction or depolarization field. If the charges of A and B ions at two zigzag edges of nanoribbons are reduced to δ/2 and −δ/2 with other ions unchanged, the total dipole moment and areal polarization will be respectively -1/2 nm δd and $-\delta/6\sqrt{3}d$.

Such polarizations also emerge in 1D nanoribbons with two parallel edges. In most experimental designs detecting the potential difference between two surfaces, the two electrodes are generally placed parallel instead of forming an angle of 60 or 120 degree, which already breaks the $D_{3h}$ symmetry when dealing with the binary honeycomb monolayer. In those cases, the potential difference induced by the polarization can be detected by two electrodes at zigzag edges instead of armchair edges (see Fig. 1(d)).

Below we simulate a series of such unconventional ferroelectric systems, and the calculations are performed based on density-functional-theory (DFT) methods implemented in the Vienna Ab initio Simulation Package (VASP 5.4.4) code[11]. The projector augmented wave (PAW) potentials with the generalized gradient approximation (GGA) in the Perdew-Burke-Ernzerhof (PBE)[12] form are used to treat the electron-ion interactions. In particular, the van der Waals correction of Grimme with zero-damping function[13] is applied for all 2D systems, where a vacuum space of 20 Å is set in the vertical direction. The kinetic energy cutoff is set to be 500 eV, the convergence threshold for self-consistent-field iteration is set to be $10^{-6}$ eV and the atomic positions were fully optimized until the forces on all atoms are less than 0.01 eV·Å$^{-1}$. The Berry phase method[2] is used to calculate the polarization, and the solid state nudged-elastic-band SS-NEB)[14] method is adopted to calculate the ferroelectric switching pathway.

The polarization of ferroelectrics should be switchable, which will be challenging for most 2D binary honeycomb structures (e.g., BN monolayer with rigid covalent bondings). Although they can be deemed as arrays of polar AB zigzag chains (see Fig. 2(a)), the ion displacements of an A-B bond length (i.e., one third of the lattice constant along the armchair direction) during switching should result in high barriers. In a previous study[15] we proposed a possible ferroelastic switching pathway (equivalent to 90 degree ferroelectric switching) of a binary honeycomb monolayer with malleable metallic bondings like InBi monolayer, where the 180 degree switching can be realized by ferroelastic switching for twice with the same switching barrier of 0.15 eV/f.u. and a square lattice as the transition state. Such InBi monolayer is yet to be synthesized, while similar switching may be applicable to some prevalent 2D honeycomb materials like GaSe and InSe with $D_{3h}$ symmetry displayed in Fig. 2(b), which may clarify the recent experimental detection of in-plane ferroelectricity in GaSe[16]. As shown in Fig. 2(c), the switching barrier of such pathway with a square lattice as the intermediate state is estimated to be 0.45 eV/f.u. for InSe monolayer, which is further reduced to ~0.3 eV/f.u. in two other pathways respectively with a tetragon-octagon lattice and 1T phase as the intermediate state. Herein the paraelectric phases seem to be metastable without imaginary frequency, also distinct from conventional ferroelectricity that can be described by soft mode model.

In some cases where the AB ions are weakly bonded, the switching barrier can be further reduced by orders of magnitude, e.g., the cyanuric acid-melamine (CAM) hydrogen-bonded honeycomb network shown in Fig. 3(a). It is composed of two types of molecules with different electronegativity, which respectively carries a charge of 0.02e and -0.02e in the network. It can also be deemed as bundles of polar AB zigzag chains that may be reversed by breaking the inter-chain hydrogen bond, which requires an energy cost of only ~22 meV/atom. Such barrier can be further reduced in ABA trilayer graphene also with $D_{3h}$ symmetry in Fig. 3(b), where the three sites marked by red, blue and green circles respectively denote the perpendicular interlayer overlapping sites of 1, 2 and 3 carbon atoms. The sequence of 3 sites with different charges can be altered during the switching via interlayer sliding,[17] crossing a barrier of ~12 meV/unitcell. If those systems are cut into nanoribbons with parallel edges breaking the $D_{3h}$ symmetry, as shown in Fig. 3(c) and (d), the estimated polarizations for armchair nanoribbons of CAM and tri-layer graphene are respectively 3.20 and 0.77 pC/m along the periodic direction, which are respectively reduced to 0.64 and 0.20 pC/m between two edges for corresponding zigzag nanoribbons due to the depolarization fields.

Another typical case in point is the zinc-blende (ZB) structures with space group F-43m. In a ZB thin film, the two parallel surfaces will break such crystal symmetry and give rise a vertical polarization. For an ideal model of ZB [001] thin-film composed with A site cations with charge $\delta$ and B site anions with charge $-\delta$, the dipole moment along [001] direction is estimated to be $\delta nm|a|/2$ by the summation of $\Delta P = \frac{1}{\Omega} \int r\delta(r) d^3r$, where n, m, |a| denote the number of units along the zigzag and armchair directions, lattice constant along the polarization direction, respectively. Over the total volume $nm|a|^3/2$, the average polarization of $\delta/|a|^2$ is also constant and does not depend on thickness, which should maintain in macroscale although the value actually measured in experiments can be influenced by edge reconstructions.

For most ZB structures with rigid and brittle covalent bondings like SiC, the switching of such polarization will usually give rise to ultrahigh barriers upon ion displacements of half of the lattice constant |a|. In previous DFT studies on ZB silver monohalides[6,8] like AgI, feasible switching pathways have been proposed with barriers below 0.15 eV/f.u.. The polarization of such monovalent system is estimated to be 40 μC/cm$^2$ (corresponding to half of the quantum $e|a|/\Omega_0$), and should be further enhanced in divalent systems like MgSe,[18] which is estimated to be 91 μC/cm² (5.93 $e\cdot$Å per unitcell, equals to a quantum $e|a|/\Omega_0$) according to the polarization evolution in Fig. 4(b). A moderate barrier can also be ensured by the covalent-like ionic bondings due to the long-range Coulomb interactions,[19] which is revealed to be 0.18 eV/f.u. according to the switching pathway, where the MgSe crystal is deemed as bundles of switchable polar zigzag MgSe chains marked by the red circles in Fig. 4(b).

To maintain a perfect stoichiometric ratio in an ideal slab model, the two surfaces of a MgSe [001] thin-film are respectively terminated by Mg and Se. The bi-stable states in Fig. 5(a) are slabs with same thickness cut from the same bulk phase along the same orientation, while their polarizations are aligned in opposite directions. The reversible polarization is estimated to be 0.33 $e\cdot$Å per unitcell, much lower compared with the bulk phase. However, it leads to a high electrostatic potential energy difference over 3.2 eV between two surfaces. A vertical polarization also emerges in the [111] thin-film displayed in Fig. 5(b), with a similar value of 0.33 $e\cdot$Å per unitcell, leading to an even higher electrostatic potential energy difference over 3.8 eV between two surfaces.

Such ferroelectric model may clarify the previously detected mysterious ferroelectricity in ZB structures of $Zn_xCd_{1-x}Te$[9]. It is also noteworthy that long ion displacements in the above-mentioned systems are equally important to the formation of unconventional ferroelectricity as the breaking of crystal symmetry by the edges. In the conventional model of ferroelectrics proposed by Abrahams, no atom in the unitcell should be displaced more than about 1 Å along the polar direction.[20] In the above-mentioned systems, however, the long-ion displacements over a bond length with moderate barriers make those polarizations switchable. Such ion displacements can be even over a lattice constant in some ferroelectric ion conductors with ion vacancies.[7]

In summary, we show the existence of unconventional ferroelectricity in some crystals that do not belong to previously acknowledged ferroelectric point groups, and propose two factors that have long been ignored in previous studies: 1. The edges of crystals are not taken into consideration by the Neumann's principle, while herein the electric polarizations even in macroscale are highly dependent on boundaries, which usually break the symmetry of those non-polar crystals. 2. The ion displacements of ferroelectrics are supposed to be less than 1 Å in Abrahams' conditions, while long ion displacements with moderate barriers are enabled in the above systems and facilitate such unconventional ferroelectric switching. Such breaking of traditional rules should significantly expand the range of available ferroelectrics for study, which are to be further explored both theoretically and experimentally.

Acknowledgement

This work is supported by the National Natural Science Foundation of China (Nos. 22073034).

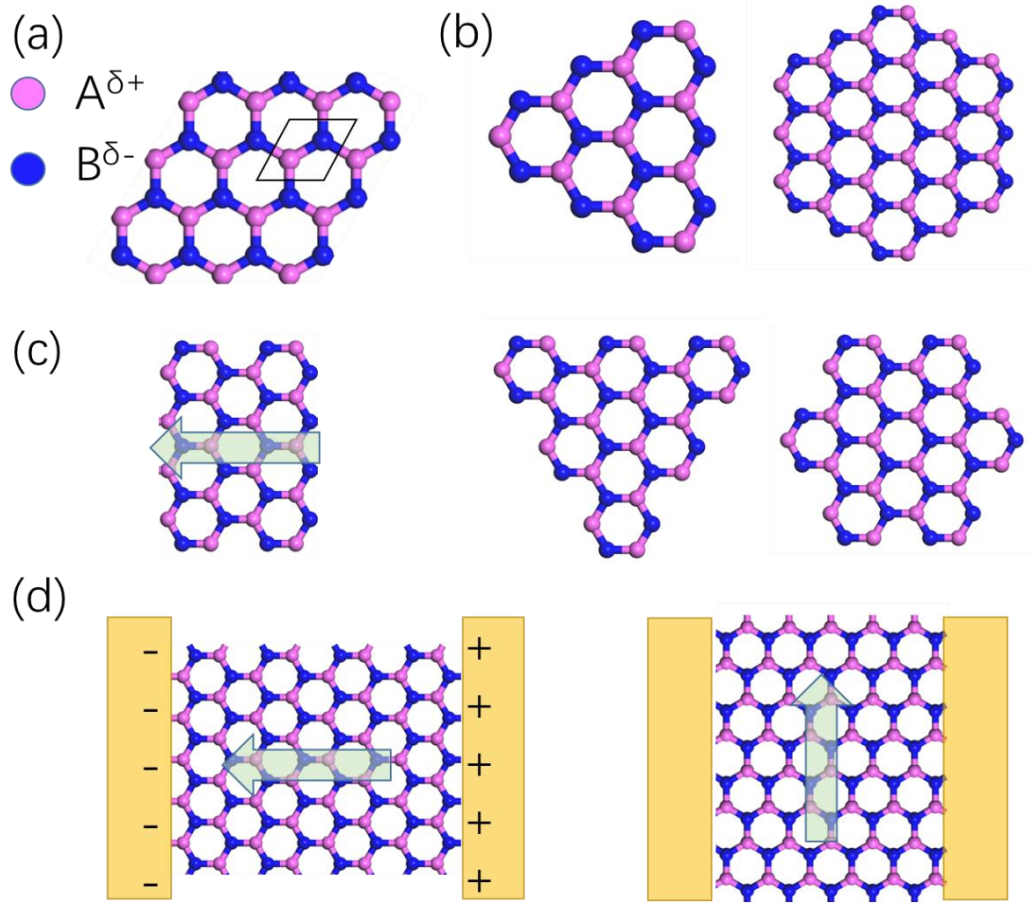

FIG. 1. Geometric structures of (a) honeycomb lattice composed of $A^{\delta+}$ and $B^{\delta-}$, (b) non-polar AB flakes with edges that do not break the $D_{3h}$ symmetry, (c) polar rectangular AB flake, (d) zigzag and armchair AB nanoribbons, where the polarization of the former one can be detected by two parallel electrodes in yellow at the edges. The green arrows denote the polarization directions.

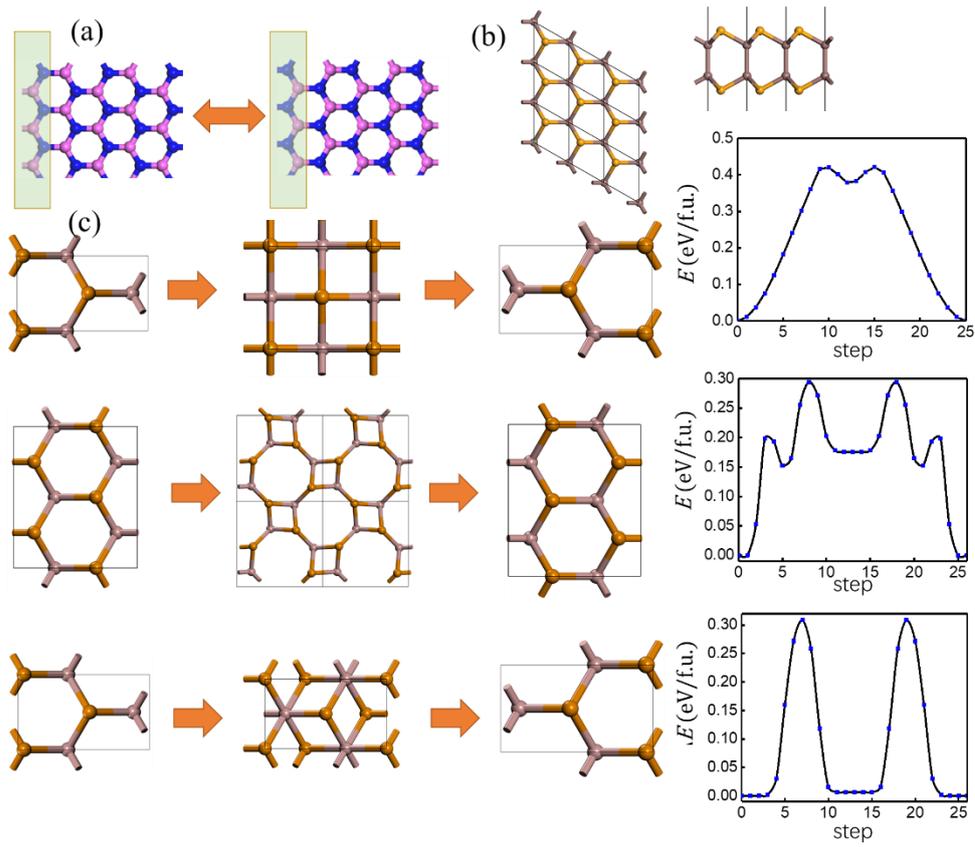

FIG. 2. (a) Ferroelectric switching of 2D AB honeycomb monolayer. (b) Geometric structure of InSe monolayer (overview and sideview) and (c) its three possible ferroelectric switching pathways with different intermediate states.

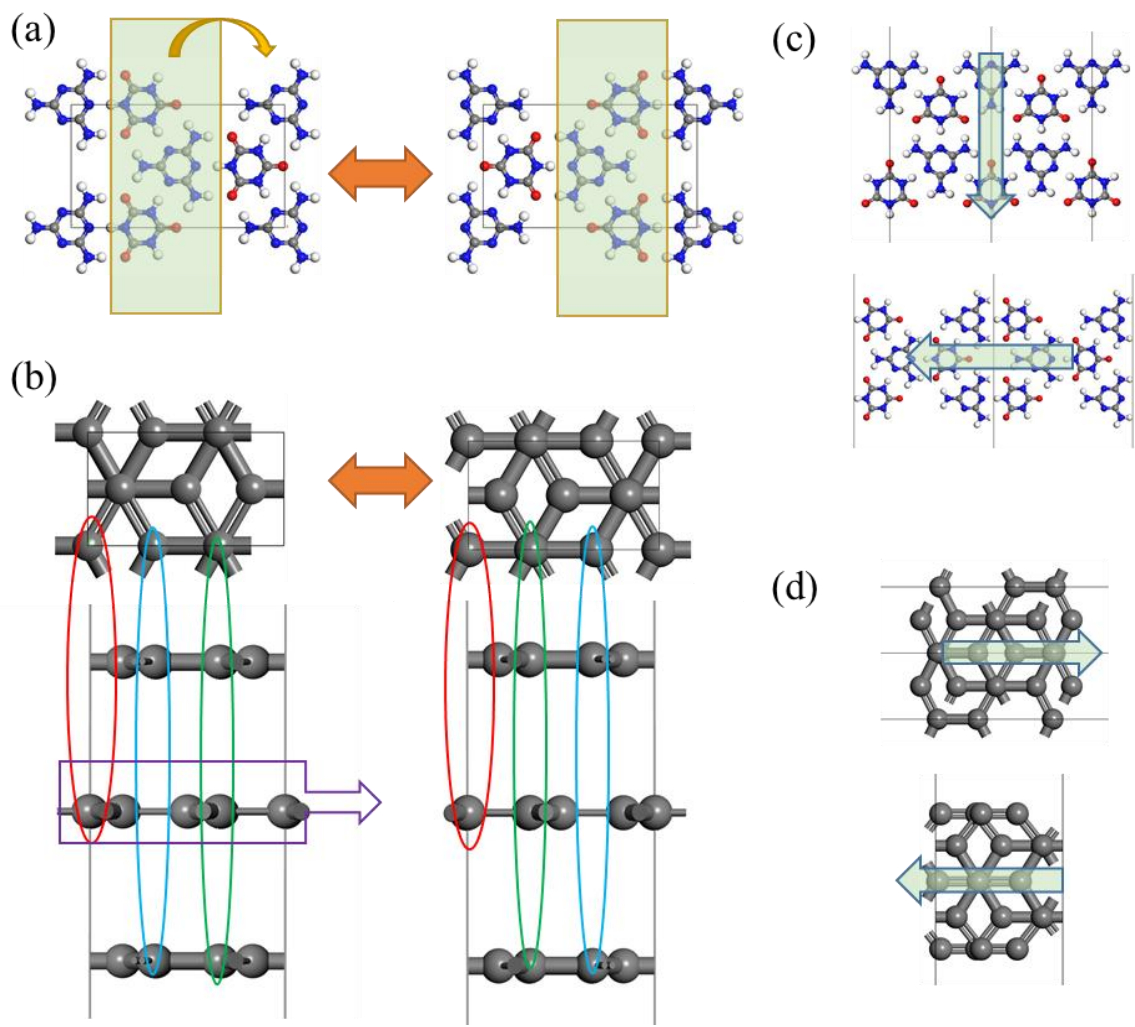

FIG. 3. Ferroelectric switching of (a) CAM and (b) ABA stacking trilayer graphene. The formation of in-plane polarizations in the armchair and zigzag nanoribbons of (c) CAM and (d) ABA stacking trilayer graphene. The green arrows denote the polarization directions.

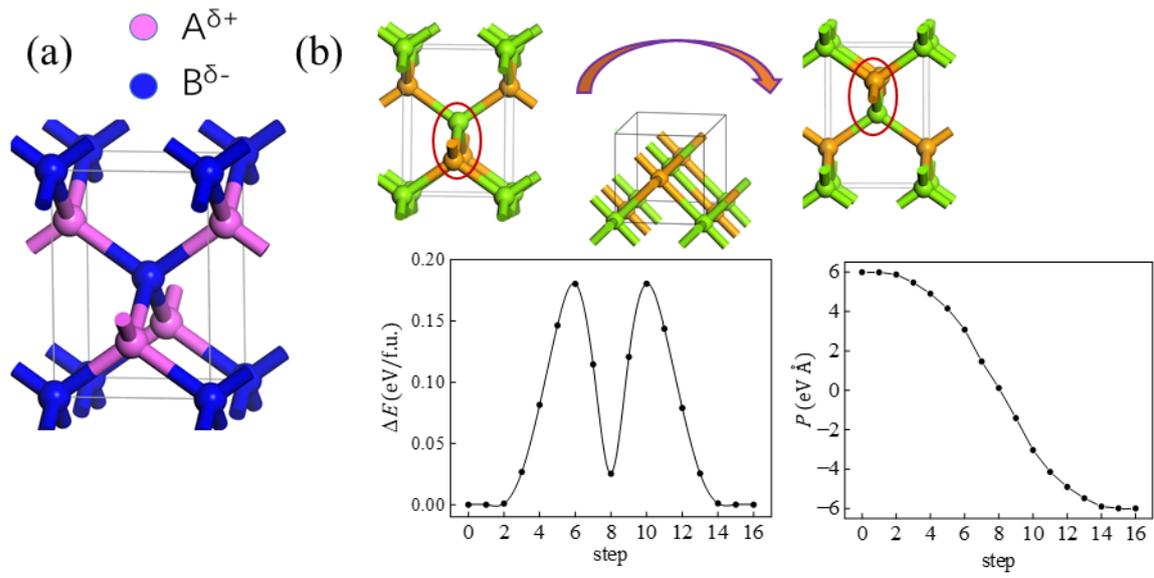

FIG. 4. (a) ZB structure composed of $A^{\delta+}$ and $B^{\delta-}$ ions. (b) The switching pathway and polarization evolution of ZB MgSe, where the evolution of a zigzag MgSe chain is marked in red circles.

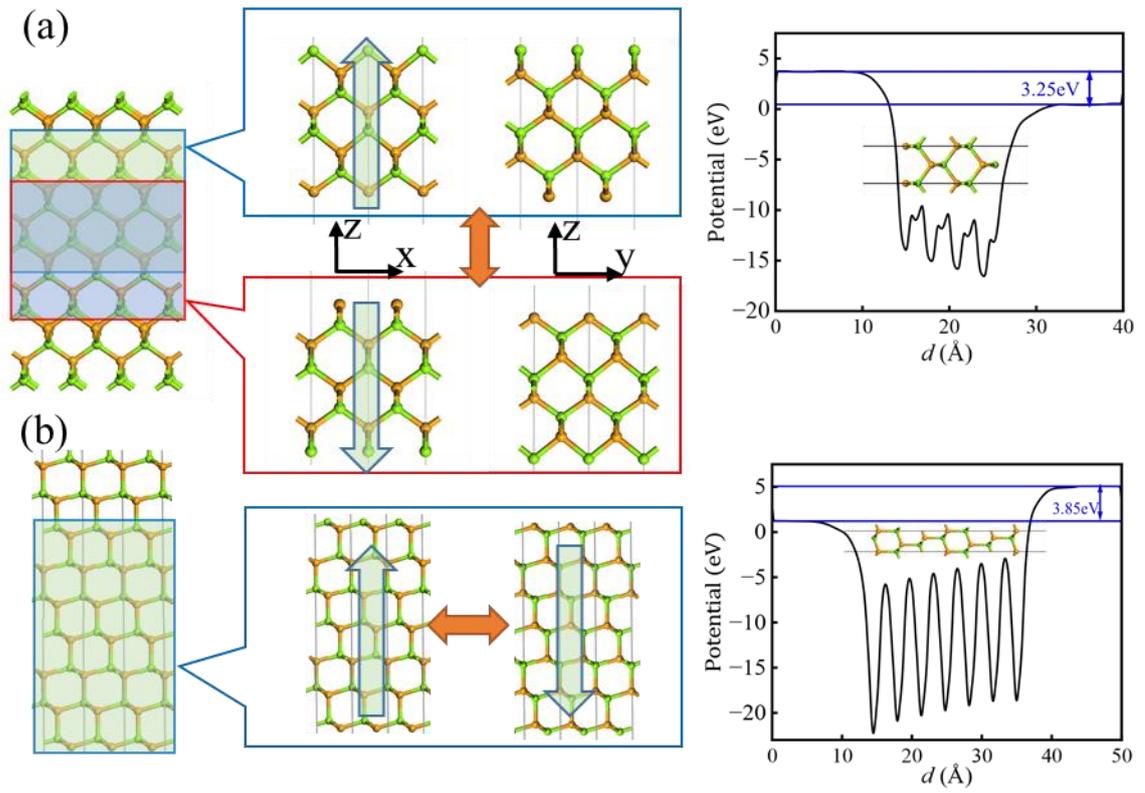

FIG. 5. Geometric structures and the calculated planar-average electrostatic potentials for few-layer MgSe (a) [001] and (b) [111] thin films.